\documentclass[a4paper,11pt]{article}

\usepackage{jheppub}
\usepackage{bm}
\usepackage{color}
\usepackage[usenames,dvipsnames,svgnames,table]{xcolor}
\usepackage{amsmath}
\usepackage{amssymb}
\usepackage{graphicx}
\usepackage{slashed}
\usepackage{soul}
\usepackage{multirow}

\def\beq{\begin{equation}}
\def\eeq{\end{equation}}
\def\bea{\begin{eqnarray}}
\def\eea{\end{eqnarray}}
\def\ben{\begin{enumerate}}
\def\een{\end{enumerate}}

\def\lsim{\mathrel{\raise.3ex\hbox{$<$\kern-.75em\lower1ex\hbox{$\sim$}}}}
\def\gsim{\mathrel{\raise.3ex\hbox{$>$\kern-.75em\lower1ex\hbox{$\sim$}}}}
\def\ifmath#1{\relax\ifmmode #1\else $#1$\fi}

\def\simlt{\stackrel{<}{{}_\sim}}
\def\simgt{\stackrel{>}{{}_\sim}}

\title{Supersymmetry and LHC Missing Energy Signals}

\author[a,b]{Marcela~Carena,}
\author[c]{James Osborne,}
\author[c]{Nausheen~R.~Shah,}
\author[b,d]{Carlos~E.~M.~Wagner}

\affiliation[a]{Fermi National Accelerator Laboratory, P.~O.~Box 500, Batavia, IL 60510, USA}
\affiliation[b]{Enrico Fermi Institute and Kavli Institute for Cosmological Physics, University of Chicago, Chicago, IL 60637, USA}
\affiliation[c]{Department of Physics \& Astronomy, Wayne State University, Detroit, MI 48201, USA}
\affiliation[d]{HEP Division, Argonne National Laboratory, 9700 Cass Ave., Argonne, IL 60439, USA}

\emailAdd{carena@fnal.gov}
\emailAdd{jaosborne@wayne.edu}
\emailAdd{nausheen.shah@wayne.edu}
\emailAdd{cwagner@anl.gov}

\preprint{FERMILAB-PUB-18-506-T
\\\phantom{0} \hfill EFI-18-15
\\\phantom{0} \hfill WSU-HEP-1806}

\abstract{
Current analyses of the LHC data put stringent bounds on strongly interacting supersymmetric particles, restricting the masses of squarks and gluinos to be above the TeV scale. However, the supersymmetric electroweak sector is poorly constrained. In this article we explore the consistency of possible LHC missing energy signals with the broader phenomenological structure of the electroweak sector in low energy supersymmetry models. As an example, we focus on the newly developed Recursive Jigsaw Reconstruction analysis by ATLAS, which reports interesting event excesses in channels containing di-lepton and tri-lepton final states plus missing energy. We show that it is not difficult to obtain compatibility of these LHC data with the observed dark matter relic density, the bounds from dark matter direct detection experiments, and the measured anomalous magnetic moment of the muon. We provide analytical expressions which can be used to understand the range of gaugino masses, the value of the Higgsino mass parameter, the heavy Higgs spectrum, the ratio of the Higgs vacuum expectation values $\tan \beta$, and the slepton spectrum obtained in our numerical analysis of these observables.
}

\begin{document}
\maketitle
\flushbottom

\section{Introduction}
\label{sec:introduction}

The Minimal Supersymmetric Standard Model (MSSM)~\cite{Nilles:1983ge,Haber:1984rc,Martin:1997ns} provides a well defined extension of the Standard Model (SM). Supersymmetric particles come in the same super-multiplets as the SM particles and hence carry the same quantum numbers as them under the SM gauge groups. Moreover, the dimensionless couplings of the SM particles with the supersymmetric partners are identified with the well known gauge and Yukawa couplings of the standard particles. However, the masses of the supersymmetric particles are determined by supersymmetry breaking parameters that are a priori unknown and should therefore be measured experimentally. 

Searches at the LEP experiment have put bounds on most supersymmetric particle masses to be above about 100~GeV (see, for example, Refs.~\cite{Barate:2000tu,Abdallah:2003xe}). A notable exception is the bino, the supersymmetric partner of the hypercharge gauge boson, which, due to its lack of couplings to the gauge bosons, could not be easily constrained by measurements at this electron-positron collider~\cite{Dreiner:2007fw}. Information on the masses of supersymmetric particles can also be obtained indirectly by flavor physics and by precision measurements. The lack of flavor changing neutral currents implies that scalar supersymmetric mass parameters must be flavor diagonal or be much larger than a~TeV~\cite{Gabbiani:1996hi,Martin:1997ns}. $B$-physics observables may still be subject to large corrections and, as we shall discuss, can lead to constraints on the charged Higgs as well as the stop and chargino masses. On the other hand, due to fast decoupling properties, precision electroweak observables put only moderate constraints on the supersymmetry breaking parameters~\cite{Pierce:1996zz}.

The LHC is currently setting stringent constraints on the strongly interacting sparticles~\cite{Aaboud:2017vwy,Sirunyan:2018vjp,Aaboud:2017hrg,Aaboud:2017aeu,Sirunyan:2018omt,Sirunyan:2018lul}. The current bound on the gluino is above a TeV, independent of its decay modes and the mass differences with the other sparticles. Assuming they are approximately degenerate in mass, similar bounds exists on the first and second generation squarks. Bounds on sbottoms are somewhat weaker, whereas the stop masses are bounded to be above about 500~GeV independent of their decay modes. Moreover, the measured value of the Higgs mass demands the average mass of the stops to be $\mathcal{O}(\gtrsim$ the TeV scale)~\cite{Carena:2011aa,Hahn:2013ria,Draper:2013oza,Bagnaschi:2014rsa,Vega:2015fna,Lee:2015uza,Bahl:2017aev}.

Much less is known about the weakly interacting particles, including the superpartners of the hypercharge and weak gauge bosons, the lepton scalar partners, and the heavy Higgs sector. The LHC is starting to be sensitive to these particles and it is expected that by the end of the high luminosity LHC run, the masses of these particles will be probed for values significantly above the LEP bounds.

Recently, the ATLAS collaboration has reported an excess of di-lepton and tri-lepton plus missing energy events~\cite{Aaboud:2018sua}. Such final state events are expected to provide the main probes of the production of charginos and neutralinos at hadron colliders~\cite{Baer:1994nr}. This excess of events has been obtained by using a newly developed Recursive Jigsaw Reconstruction (RJR) method~\cite{Jackson:2016mfb,Jackson:2017gcy}, where the missing energy requirements are found to be less severe while maintaining an estimated signal sensitivity comparable to that one of more conventional searches. Moreover, the background in these searches was determined by data driven methods and the overlap of the signal regions with the ones in conventional searches is small. Hence, it is possible that these searches could be sensitive to a region of parameters that lead to no apparent signal events in conventional searches~\cite{Aaboud:2018jiw,Sirunyan:2018ubx}. Furthermore, as we shall discuss in more detail, the GAMBIT collaboration~\cite{Kvellestad:2018akf,Athron:2018vxy} has recently argued that the exclusion limits obtained by the ATLAS and CMS experiments in conventional searches may be relaxed due to the differences between the signatures associated with the simplified model analyzed by the ATLAS and CMS collaborations and those associated with the full low energy supersymmetry model. 

Although the excess of events reported by ATLAS cannot be taken as a compelling signal of new physics at this point, we use it as a guide to study the current constraints on the electroweak sector of the MSSM. In particular, by assuming the event excesses reported by the ATLAS collaboration to be a signal of new physics, we determine the necessary MSSM gaugino sector leading to an explanation of the observed signatures. We then combine this information with that provided by the observed dark matter (DM) relic density, the current bounds from DM direct and indirect detection experiments, and the measured value of the anomalous magnetic moment of the muon $a_\mu$. Using all this information we can determine the range of allowed values not only for the gaugino masses~($M_1$ and $M_2$), but also for the Higgsino mass parameter~($\mu$), the ratio of Higgs vacuum expectation values~($\tan \beta$), and the heavy Higgs and slepton mass spectrum. This work shows that all these experimental constraints can be easily satisfied, with the obtention of the DM relic density (assuming it to have a thermal origin) providing the most relevant constraint. We provide analytical expressions that allow an understanding of the dependence of the observables on the supersymmetric mass parameters, as well as a numerical study of the allowed electroweak sector parameter space.

This article is organized as follows. In Sec.~\ref{sec:ATLASexcesses} we shall discuss the possible interpretation of the ATLAS event excesses. In Sec.~\ref{sec:DarkMatter} we shall review the constraints coming from the observed DM relic density and the lack of direct detection of DM interacting with nuclei. We also comment on the indirect detection prospects for this scenario. In Sec.~\ref{sec:g-2} we shall analyze the constraints coming from the measurement of the anomalous magnetic moment of the muon. We briefly comment on flavor observables which may be related to our scenario in Sec.~\ref{sec:flav}. Finally we present a study of the variation of the main observables with the parameters of the model together with a benchmark scenario in Sec.~\ref{sec:Explorations}, and reserve Sec.~\ref{sec:Conclusions} for our conclusions.

\section{ATLAS Event Excesses}
\label{sec:ATLASexcesses}

The ATLAS collaboration has recently released a search for chargino-neutralino production in two and three lepton final states employing RJR techniques that target specific event topologies~\cite{Aaboud:2018sua}. The analysis was performed on data corresponding to a total integrated luminosity of $36.1~\textrm{fb}^{-1}$, and finds excesses of observed events above the estimated background with 2--3$\sigma$ significance in several signal regions (SRs). This analysis searches for chargino-neutralino pair production, assuming a wino-like production mechanism, and decays with 100\% branching ratio into a bino-like lightest supersymmetric particle~(LSP) via on-shell $W$ and $Z$ gauge bosons. The excess of events appear in the SRs targeting the low mass, $m_{\widetilde{\chi}_1^\pm/\widetilde{\chi}_2^0} \lesssim 200$~GeV, and low mass-splitting, $\Delta m \equiv m_{\widetilde{\chi}_1^\pm/\widetilde{\chi}_2^0} - m_{\widetilde{\chi}_1^0} \sim 100$~GeV, region of parameter space. In addition, these low mass splittings kinematically suppress the decay of the second lightest neutralino into a SM-like Higgs and the lightest neutralino, $(\widetilde{\chi}_2^0 \to h \ \widetilde{\chi}_1^0)$, independently of the neutralino compositions. 

\begin{figure}[t]
  \centering
  \includegraphics[width=0.4\textwidth]{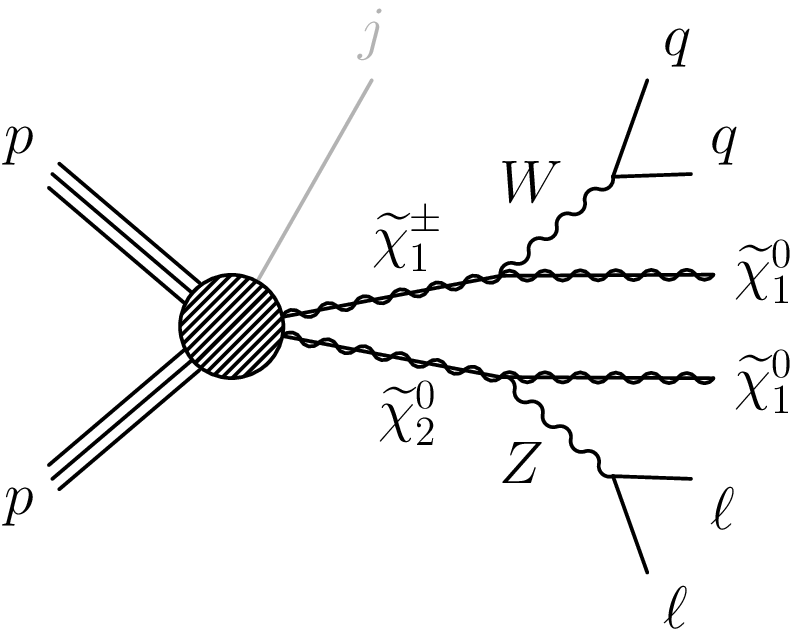}
  \hspace*{0.1\textwidth}
  \includegraphics[width=0.4\textwidth]{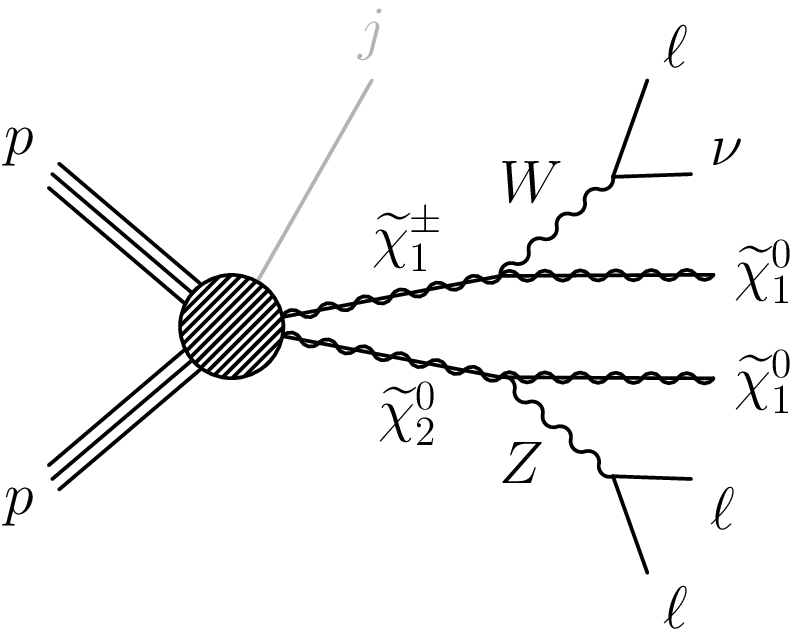}
  \caption{Feynman diagrams for the scenarios targeted by the ATLAS RJR analysis, leading to final states of either two (left) or three (right) leptons plus missing energy. Several of the SRs require an additional ISR jet, shown in light gray.}
  \label{fig:feynman_2l3l}
\end{figure}

\begin{table}[tbh]
  \centering
  \renewcommand{\arraystretch}{1.2}
  \begin{tabular}{l c c c c}
    \hline
    Signal Region & Observed Events & BG Events & Events above BG & Significance ($Z$) \\
    \hline \hline
    SR2$\ell_\textrm{Low}$ & 19 & $8.4 \pm 5.8$ & $10.6 \pm 5.8$ & 1.39 \\
    SR2$\ell_\textrm{ISR}$ & 11 & $2.7^{+2.8}_{-2.7}$ & $8.3^{+2.8}_{-2.7}$ & 1.99 \\
    SR3$\ell_\textrm{Low}$ & 20 & $10 \pm 2$ & $10 \pm 2$ & 2.13 \\
    SR3$\ell_\textrm{ISR}$ & 12 & $3.9 \pm 1.0$ & $8.1 \pm 1.0$ & 3.02 \\
    \hline
  \end{tabular}
  \caption{Expected and observed events for the 2$\ell$ and 3$\ell$ SRs, as well as the significance of the excess ($Z$). The number of observed events, background estimates and significance of the excess are taken from Ref.~\cite{Aaboud:2018sua}. The errors on the background show statistical plus systematic uncertainties. The third column has been added to show the estimated number of events above expected background.}
  \label{tab:events_sig}
\end{table}

\begin{figure}[bh]
  \centering
  \includegraphics[width=0.65\textwidth]{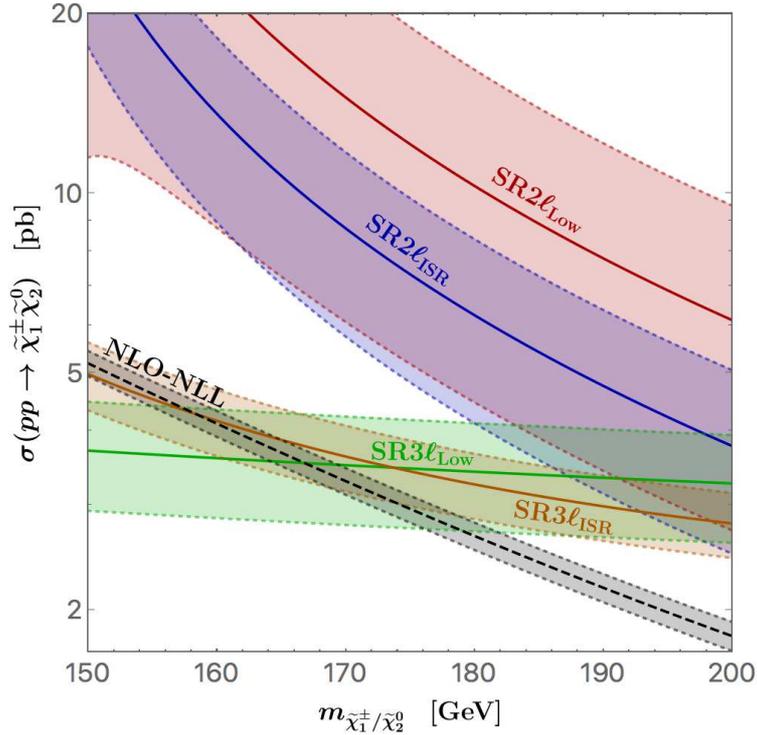}
  \caption{Signal cross sections that reproduce the observed excesses in each SR as a function of $m_{\chi^\pm/\chi_2^0}$, assuming $\Delta m = 100 $ GeV, with $\pm 1 \sigma$ bands obtained by propagating the background uncertainties. The black dashed line denotes the NLO-NLL pure wino-like $\widetilde{\chi}_1^\pm \widetilde{\chi}_2^0$ production cross section with a $\pm 1 \sigma$ uncertainty band.}
  \label{fig:sig_xsecs}
\end{figure}

The topologies considered in the analysis are shown in Fig.~\ref{fig:feynman_2l3l}. The four low mass SRs are labeled by the number of final state leptons and whether the search targets events with a hard ISR jet: SR2(3)$\ell_\textrm{ISR}$ require an ISR jet whereas SR2(3)$\ell_\textrm{Low}$ do not. The excess number of events as well as the significance of the excess in each signal region is summarized in Table~\ref{tab:events_sig}. To estimate the signal cross section, $\sigma(pp \rightarrow \widetilde{\chi}_1^\pm \widetilde{\chi}_2^0)$, required to produce each excess, we used the binned grids of acceptances and efficiencies for each SR as provided by the ATLAS collaboration~\cite{RJR_aux} together with a linear interpolation between bins, assuming a mass difference $\Delta m = 100$~GeV. The results are shown in Fig.~\ref{fig:sig_xsecs} as solid lines. The bands show the $\pm 1 \sigma$ uncertainties estimated by propagating the background uncertainties. For reference, we also show the NLO-NLL wino-like $\widetilde{\chi}_1^\pm \widetilde{\chi}_2^0$ production cross section~(black dashed line) with a $\pm 1 \sigma$ uncertainty band~\cite{LHC_XSWG,Fuks:2012qx,Fuks:2013vua}. The production cross-section of Higgsino-like $\widetilde{\chi}_1^\pm \widetilde{\chi}_2^0$~(not shown) is approximately a factor of 4 smaller. In the MSSM, generically the neutralinos are expected to be admixtures rather than pure states. As such, the pure wino cross section denoted in Fig.~\ref{fig:sig_xsecs} should be treated as an upper bound for $\widetilde{\chi}_1^\pm \widetilde{\chi}_2^0$ production in the MSSM, and may be significantly reduced if $|\mu|$ is of the same order as $|M_2|$. 

As can be seen from Fig.~\ref{fig:sig_xsecs}, the upper region in the low mass window, $(m_{\widetilde{\chi}_1^\pm/\widetilde{\chi}_2^0},\, m_{\widetilde{\chi}_1^0}) \sim (200, 100)$~GeV, is found to be preferred in order to accommodate all observed excesses. However, for a 200 GeV wino-like chargino-neutralino pair, the NLO+NLL production cross section~\cite{LHC_XSWG,Fuks:2012qx,Fuks:2013vua} is estimated to be only $\sigma(p p \rightarrow \widetilde{\chi}_1^\pm \widetilde{\chi}_2^0) \approx 1.8$~pb, more than $2.5 \sigma$ below the lowest estimated signal cross section. The three lepton excesses, however, can be simultaneously accommodated by gaugino masses of approximately 165 GeV, where the wino-like production cross section is estimated to be $\sim 3.6$~pb. Hence for this low mass region, the production of mostly wino-like $\widetilde{\chi}_1^\pm/\widetilde{\chi}_2^0$ with small Higgsino components can easily be consistent with the excess. For the remainder of this article we focus on the three lepton excesses, which have the greatest significance. 

Recently, the GAMBIT collaboration reinterpreted these searches~\cite{Kvellestad:2018akf,Athron:2018vxy}, including the effect of the production of the heavy charginos and neutralinos, $\widetilde{\chi}_{3,4}^0,\widetilde{\chi}_2^\pm$, which are mostly Higgsino-like in our scenario. They concluded that due to the differences between the signatures of the full model and the simplified model analyzed by the ATLAS and CMS collaborations, the limits obtained through these experimental analyses, e.g. those obtained in Ref.~\cite{Aaboud:2018jiw,Sirunyan:2018ubx}, are not always applicable to the full MSSM model. They also provide likelihood contours from collider data, including not only the ATLAS RJR searches for charginos and neutralinos but also all other conventional searches for charginos and neutralinos at ATLAS and CMS. They showed that regions with chargino masses of about 150~GeV and lightest neutralino masses of about 50~GeV provide the best fit to the collider data. Also preferred are values of $|\mu|$ of about 250~GeV. The region of chargino masses of about 200~GeV and lightest neutralino masses of 100~GeV also appears to be among the preferred regions in their fits, as can be understood from our discussion above. 

 As discussed in the introduction, we shall take the interpretation of these signals as a guidance for the determination of the parameter space of the gaugino and Higgsino mass parameters, in order to study the current experimental constraints on the electroweak sector of low energy supersymmetry models. For the electroweakino sector, although we shall concentrate on the region with chargino masses close to 150~GeV and lightest neutralino masses close to 50~GeV, we shall comment on the phenomenology of heavier chargino and neutralino masses.

\section{Dark Matter Phenomenology} 
\label{sec:DarkMatter}

The possibility of explaining the observed DM relic density by the presence of a stable LSP is an interesting implication of low energy supersymmetry~\cite{Jungman:1995df,Drees:1992am}. As discussed in the previous section, the signal we are considering shows a preference for a light gaugino sector, with the heavier weak gauginos~($\widetilde{\chi}_2^0,~ \widetilde{\chi}_1^\pm$) mostly wino-like with masses of the order of 160~GeV, to allow for the large value of the production cross sections necessary to explain the excess of events observed by the ATLAS experiment. Moreover, these gauginos should decay into on-shell weak gauge bosons and the lightest bino-like neutralino, implying that the lightest neutralino should be lighter than about 70~GeV. 

This range of masses for the lightest bino-like neutralino excludes the possibility of co-annihilation with other supersymmetric particles, since either such additional particles are excluded experimentally or their presence would modify the collider signatures. Hence the only natural way of getting the proper relic density is either through resonant annihilation via either the SM-like Higgs or the $Z$ gauge boson, which imply masses for the neutralino close to 60 and 45 GeV, respectively, or via the $t$-channel interchange of light-sfermions. Since the bino couples to the Higgs and the $Z$ only via its Higgsino components~[see Eqs.~(\ref{eq:hNN}) and~(\ref{eq:Zxx}) below], $|\mu|$ is required to be of the order of a few hundred GeV. For the case of resonant annihilation via the $Z$ gauge boson the neutralino is light enough to allow for unsuppressed decays of the SM-like Higgs bosons into pairs of these particles. However, it turns out that at large values of $\tan \beta$ the coupling of the Higgs to the lightest neutralino is sufficiently small as to prevent a significant branching ratio for this decay mode, even in the case of values of $|\mu|$ close to the weak scale.

Regarding the $t$-channel annihilation possibility, since we are assuming all squarks to be heavy, the only possibility is associated with light sleptons. Given the LHC direct search bounds on sleptons~\cite{Aaboud:2017leg,Aaboud:2018jiw}, this possibility can only be realized for light staus~\cite{Sirunyan:2018vig}. If the lightest neutralino has some Higgsino mixing and for large values of $\tan\beta$, for which the $\tau$ Yukawa coupling is enhanced, a consistent relic density may be obtained. In this case, the amplitude for the neutralino annihilation is proportional to~\cite{pierce:decomp}
\begin{equation}
\mathcal{A}_{\widetilde{\tau}}\left( \widetilde{\chi}_1^0 \widetilde{\chi}_1^0 \to \tau^+ \tau^- \right) \propto  \frac{\mu ~m_{\widetilde{\chi}_1^0}}{\left(\mu^2 - m_{\widetilde{\chi}_1^0}^2\right)}\frac{ m_Z^2 s_W^2 m_\tau \tan\beta}{v^2 \left(m_{\widetilde{\tau}_R}^2+ m_{\widetilde{\chi}_1^0}^2\right)} \, .
\end{equation}

In general, we shall work in the large $\tan\beta$ regime for two reasons: first, it is easier to accommodate the observed Higgs mass for stop masses at the TeV scale~\cite{Carena:2011aa,Hahn:2013ria,Draper:2013oza,Bagnaschi:2014rsa,Vega:2015fna,Lee:2015uza,Bahl:2017aev}. Second, as we shall discuss in the next section, it is easier to accommodate the observed value of the anomalous magnetic moment of the muon in this regime. 

We also comment briefly on the indirect detection prospects for the scenario under consideration. Both the Higgs and $Z$ mediated annihilations are $p$-wave suppressed, and hence will not give rise to any indirect signals. On the other hand, the $t$-channel exchange of $\widetilde{\tau}_R$, can have a significant $s$-wave annihilation at large values of $\tan\beta$, giving rise to the possibility of current day signals into pairs of $\tau$ leptons. Such signals, though interesting, don't rule out this possibility~(see e.g. Refs. \cite{Hagiwara:2013qya,Han:2014nba,Ackermann:2015zua}). There is also the possibility of additional channels contributing to the relic density, as is the case in the next to minimal supersymmetric standard model~(NMSSM), where the relic density may be obtained mostly due to the $s$-wave annihilation of the singlet-like pseudoscalar. In such a case, one may also obtain large signals from present day DM annihilations coming from the galactic center~(see e.g. Ref.~\cite{Cheung:2014lqa}). 

DM scattering with nuclei can provide an efficient probe for the presence of DM in our galaxy~\cite{Goodman:1984dc}. There has been an intensive experimental program looking for the direct detection of DM in the last decades. No clear signal has been found, and the strongest experimental bounds today are coming from the PandaX~\cite{Cui:2017nnn}, LUX~\cite{Akerib:2016vxi} and XENON1T~\cite{Aprile:2017iyp} experiments. A small excess of events has been found in the last round of the XENON1T experiment~\cite{Aprile:2018dbl}, which although far from being significant, could be a hint of the possible presence of DM with spin independent~(SI) scattering cross sections with nuclei of the order of the current limit. 

As is well known, the SI scattering cross section of mostly bino DM with nuclei can be easily smaller than the current limits, particularly for negative values of $\mu \times M_1$~\cite{Ellis:2000ds,PhysRevD.63.065016,Baer:2006te,Huang:2014xua,Huang:2017kdh}. In the following, we shall briefly discuss the main reason for the preference for negative values of $\mu \times M_1$ as well as the possible existence of blind spots~\cite{Cheung:BS,Huang:2014xua,Crivellin:2015bva} for SI direct DM detection within the MSSM.

The coupling of the lightest neutralino to the SM-like Higgs~($h$) and the heavy non-SM-like Higgs~($H$) in the decoupling or alignment limit are 
\begin{eqnarray}\label{eq:hNN}
  g_{h \widetilde{\chi}_1^0 \widetilde{\chi}_1^0} &=& \left ( g_1 N_{11} - g_2 N_{12} \right ) \left ( N_{13} \cos\beta - N_{14} \sin \beta \right ) \, , \\
  g_{H \widetilde{\chi}_1^0 \widetilde{\chi}_1^0} &=& \left ( g_1 N_{11} - g_2 N_{12} \right ) \left ( N_{13} \sin \beta - N_{14} \cos \beta \right ) \, ,
\end{eqnarray} 
where $g_1$ and $g_2$ are the hypercharge and weak gauge couplings, and $N_{1j}$ are the $j^{th}$ electroweak components for the lightest neutralino mass eigenstate, where $j$ = $\{1,2,3,4\}$ = $\{$bino ($\widetilde{B}$), wino ($\widetilde{W}$), down-type Higgsino ($\widetilde{H}_d$), up-type Higgsino ($\widetilde{H}_u$)$\}$. The coupling to the $Z$ gauge boson is instead given by 
\begin{equation}\label{eq:Zxx}
  g_{Z \widetilde{\chi}_1^0 \widetilde{\chi}_1^0} = \frac{g_2}{c_W} \left ( N_{13}^2 - N_{14}^2 \right ) \, ,
\end{equation}
where we have assumed the $N_{1j}$ to be real and $c_W$ is the cosine of the weak mixing angle.

Ignoring the wino component of the mostly bino-like $\widetilde{\chi}_1^0$, the neutralino components are well approximated by the following expressions~\cite{pierce:decomp, Cheung:2014lqa}:
\begin{align}
  \frac{N_{12}}{N_{11}} &\approx 0 \, , \nonumber \\
  \frac{N_{13}}{N_{11}} &= \frac{m_Z s_W \sin \beta}{\mu^2 - m_{\widetilde{\chi}_1^0}^2} \left ( \mu + \frac{m_{\widetilde{\chi}_1^0}}{\tan \beta} \right ) \, , \nonumber \\ 
  \frac{N_{14}}{N_{11}} &= - \frac{m_Z s_W \cos \beta}{\mu^2 - m_{\widetilde{\chi}_1^0}^2} \left ( \mu + m_{\widetilde{\chi}_1^0} \tan \beta \right ) \, , \label{eq:coupbino} \\ 
  N_{11} &= \left ( 1 + \frac{N_{13}^2}{N_{11}^2} + \frac{N_{14}^2}{N_{11}^2} \right )^{- \frac{1}{2}} \, ,\nonumber 
\end{align}
where $m_Z$ is the neutral gauge boson $Z$ mass, and $s_W$ is the sine of the weak mixing angle. This leads to the following couplings of the lightest neutralino to the Higgs and the $Z$ bosons:
\begin{eqnarray}\label{eq:gBH1}
  g_{h \widetilde{\chi}_1^0 \widetilde{\chi}_1^0} &=& \frac{2 m_Z^2 s_W^2 N_{11}^2 }{v \left(\mu^2 - m_{\widetilde{\chi}_1^0}^2 \right )} \left ( m_{\widetilde{\chi}_1^0} + \mu \sin 2 \beta \right ) \, , \nonumber \\
  g_{H \widetilde{\chi}_1^0 \widetilde{\chi}_1^0} &=& -\frac{2 m_Z^2 s_W^2 N_{11}^2}{v \left ( \mu^2 - m_{\widetilde{\chi}_1^0}^2 \right )} \mu \cos 2 \beta \, , \nonumber \\
  g_{Z \widetilde{\chi}_1^0 \widetilde{\chi}_1^0} &=& - \frac{2 m_Z^3 s_W^2 N_{11}^2}{ v \left ( \mu^2 - m_{\widetilde{\chi}_1^0}^2 \right )}\cos 2 \beta \, ,
\end{eqnarray}
where $v=246$~GeV. 

The coupling of the Higgs bosons to up and down quarks are given by
\begin{align}
  g_{ddh} &= \frac{m_d \sqrt{2}}{v} \, , \\
  g_{uuh} &= \frac{m_u \sqrt{2}}{v} \, , \\
  g_{ddH} &= -\frac{m_d\sqrt{2} \tan \beta}{v} \, , \\
  g_{uuH} &= \frac{m_u\sqrt{2} \tan \beta}{v} \, ,
\end{align}
where $m_u$ and $m_d$ are the up and down quark masses. In the above, we have ignored the finite corrections to the Higgs couplings coming from the decoupling of squarks and gluinos~\cite{Hempfling:1993kv,Hall:1993gn,Carena:1994bv,Carena:1998gk,Carena:1999bh} since they are small in the region of parameters we are interested in, where $|\mu|$ is much smaller than the squark and gluino masses. 

In the region of parameters we are investigating, the cross section for SI direct detection is controlled predominantly by the exchange of the Higgs bosons. Also including the approximate contributions due to heavy squarks and taking the limit $m_{\widetilde{\chi}_1^0}^2 \ll \mu^2$ for a predominantly bino-like LSP, the SI cross section for the scattering of DM off protons is given by~(similar expression holds for scattering off neutrons)~\cite{Huang:2014xua, Cheung:2014lqa, Crivellin:2015bva}
\begin{eqnarray}
  \sigma_p^\textrm{SI} &\simeq& \frac{4 m_Z^4 s_W^4 m_p^2 m_r^2}{\pi v^4 \mu^4} N_{11}^4 \left [ -\left ( F_{d}^{(p)} + F_{u}^{(p)} \right ) \frac{( m_{\widetilde{\chi}_1} + \mu \sin 2 \beta )}{m_h^2} \right . \nonumber \\
  && \left . - \left ( - F_{d}^{(p)} + \frac{F_u^{(p)}}{\tan^2 \beta} \right ) \frac{\mu \tan \beta \cos 2 \beta}{m_H^2} - \frac{F_{u}^{(p)} \left ( m_{\widetilde{\chi}_1^0} + \mu / \tan \beta \right ) + F_{d}^{(p)} \left(m_{\widetilde{\chi}_1^0} + \mu \tan \beta \right )}{2 m_{\widetilde{Q}}^2} \right ]^2 \, , \nonumber \\
  \label{eq:sig}
\end{eqnarray}
with $F_{u}^{(p)} \equiv f_{u}^{(p)} + 2 \times \frac{2}{27} f_{TG}^{(p)} \approx 0.15$ and $F_{d}^{(p)} = f_{Td}^{(p)} + f_{Ts}^{(p)} + \frac{2}{27} f_{TG}^{(p)} \approx 0.14$, $m_p$ is the proton mass, $m_r = m_p m_{\widetilde{\chi}_1^0}/(m_p + m_{\widetilde{\chi}_1^0})$ is the reduced mass, and $m_{\widetilde{Q}}$ is the common squark mass. Since $F_{u}^{(p)} \approx F_{d}^{(p)}$, in the large $\tan \beta$ limit this expression becomes proportional to
\begin{equation}
  \sigma_p^\textrm{SI} \propto \frac{m_Z^4}{\mu^4} \left [ 2 (m_{\widetilde{\chi}_1^0}+2 \mu / \tan \beta )\frac{1}{m_h^2} + \mu \tan \beta \frac{1}{m_H^2} + (m_{\widetilde{\chi}_1^0} + \mu \tan \beta / 2) \frac{1}{m_{\widetilde{Q}}^2} \right ]^2 \, .
  \label{eq:siglargetb}
\end{equation}

It is hence clear that the cross section is reduced for negative values of $\mu \times m_{\widetilde{\chi}_1^0}$, where we shall assume $m_{\widetilde{\chi}_1^0} \simeq M_1$ to be positive, where $M_1$ is the bino mass parameter. Consequently, while positive values of $\mu$ tend to lead to conflict with the current bounds from the PandaX, XENON1T and LUX experiments, negative values of $\mu$ easily lead to consistency with these constraints in the large $\tan \beta$ regime. Depending on the values of the neutralino mass, the heavy Higgs boson mass, the squark masses and $\tan \beta$, the SI cross section may be close to the current bound, or may be efficiently suppressed in the proximity of blind spots that occur when~\cite{Huang:2014xua, Cheung:2014lqa, Crivellin:2015bva}
\begin{equation}
  \label{eq:blindspot}
  2 \left ( m_{\widetilde{\chi}_1^0} + 2 \frac{\mu}{\tan\beta}\right)\frac{1}{m_h^2} \simeq -\mu \tan \beta \left ( \frac{1}{m_H^2} + \frac{1}{2 m_{\widetilde{Q}}^2} \right ) \, .
\end{equation}
Finally, Eq.~(\ref{eq:siglargetb}) shows a strong dependence of the SI cross section with the value of $|\mu|$, a behavior that is related to its dependence on the square of the Higgsino components.

The spin dependent~(SD) cross section, instead, depends only on the coupling to the $Z$~\cite{Agrawal:2010fh,Freytsis:2010ne}, and hence to the difference of the squares of the up and down Higgsino components. From the expression given in Eq.~(\ref{eq:gBH1}), one can see that 
\begin{equation}\label{eq:SDmu}
  \sigma^\textrm{SD} \propto \frac{m_Z^4}{\mu^4} \cos^2 ( 2 \beta ) \, ,
\end{equation}
where we have again assumed that $\mu^2 \gg m_{\widetilde{\chi}_1^0}^2$. Hence, in the large $\tan \beta$ regime and for $|\mu|$ sufficiently large, the SD cross section is suppressed by four powers of $\mu$, without any other strong parametric suppression. This behavior should be contrasted with the SI cross section which, in spite of its overall suppression by only two powers of $\mu$, may be further suppressed due to a reduction of the neutralino coupling to the 125 GeV Higgs boson together with interference effects. As we will show, for negative values of $\mu$, and $|\mu|$ sufficiently large to avoid the SD cross section limits, the SI cross section tends to be below the current experimental bounds on this quantity. However, it can come closer to the current limits depending on the precise value of $\tan \beta$ and $m_H$.

\section{Anomalous Magnetic Moment of the Muon}
\label{sec:g-2}

The anomalous magnetic moment of the muon is a very relevant quantity since it may be measured with great precision and is sensitive to physics at the weak scale. The theoretical prediction within the SM may be divided in four main parts
\begin{equation}
  a_\mu = a_\mu^{\rm QED} + a_\mu^{\rm EW} + a_\mu^{\rm had}({\rm vac. \ pol.}) + a_\mu^{\rm had}(\gamma \times \gamma) \, ,
\end{equation}
where $a_\mu \equiv (g_\mu - 2)/2$. The first term $a_\mu^{\rm QED}$ represents the pure electromagnetic contribution, and is known with great accuracy, up to five loop order~\cite{Kinoshita:2004wi}. The second term denotes the electroweak contributions, which are known at the two-loop level, and are about $(153.6 \pm 1.) \times 10^{-11}$~\cite{Czarnecki:1995sz}. The hadronic contributions contain the largest uncertainty in the determination of $a_\mu$. While the vacuum polarization contributions can be extracted from the scattering process of $e^+ e^-$ to hadrons and are of order of $(7 \times 10^{-8}$~\cite{Kinoshita:1984it,Davier:2010nc,Hagiwara:2011af}), the so-called light by light contributions $a_\mu^{\rm had}(\gamma \times \gamma)$ cannot be related to any observable and have to be estimated theoretically. These are estimated to be about $105 \times 10^{-11}$~\cite{Prades:2009tw} and hence of the order of the electroweak contributions. 

Overall, the theoretical calculation of $a_\mu$ in the SM~\cite{Patrignani:2016xqp} differs from the result measured experimentally at the Brookhaven E821 experiment~\cite{Bennett:2006fi} by
\begin{equation}
  \delta a_\mu = a_\mu^{\rm exp} - a_\mu^{\rm theory} = 268 (63) (43) \times 10^{-11} \, ,
\end{equation}
where the errors are associated with the experimental and theoretical uncertainties, respectively. The discrepancy, of order 3.5$\sigma$, is of similar size as the electroweak contributions and hence can be potentially explained by new physics at the weak scale. The E821 experimental result will be tested by the upcoming Muon $g-2$ Experiment at Fermilab~\cite{Chapelain:2017syu}.

In the supersymmetric case the most relevant contributions are associated with the interchange of charginos and the superpartners of the neutral second generation leptons (sneutrinos)~\cite{Barbieri:1982aj,Ellis:1982by,Kosower:1983yw,Moroi:1995yh,Carena:1996qa,Czarnecki:2001pv,Feng:2001tr,Martin:2001st}. Assuming that there are no large mass hierarchies in the supersymmetric electroweak sector, one can write, approximately,
\begin{equation}\label{eq:amu}
  \delta a_\mu \simeq \frac{\alpha}{8 \pi s^2_W } \frac{m_\mu^2}{\widetilde{m}^2} { \rm Sgn }(\mu M_2) \tan \beta \, \simeq 130 \times 10^{-11} \left ( \frac{ 100 \ {\rm GeV}}{\widetilde{m}} \right)^2 {\rm Sgn}(\mu M_2) \tan\beta \, ,
\end{equation}
where $\alpha$ is the electromagnetic fine structure constant, and $\widetilde{m}$ is the characteristic mass of the weakly interacting sparticles. This implies that for $\tan\beta$ of order 10~(20), the overall weakly interacting sparticle mass scale must be of order 250~GeV~(350~GeV) in order to explain the current discrepancy between theory and experiment. 

In our work, we shall consider chargino and slepton masses that are quite different from each other and hence, it is relevant to provide an analytical understanding of the behavior of $a_\mu$ in that parameter regime. In the relevant approximation where $|\mu| \simgt 2 |M_2| \simgt 4 M_W$ and $m_{\widetilde{\nu}}^2 \simgt \mu^2$, one gets, 
\begin{equation}
\delta a_\mu \simeq - \frac{3 \alpha}{4 \pi s^2_W}\frac{m_{\mu}^2}{m_{\widetilde{\nu}}^2} \frac{M_2 \mu \tan\beta}{\mu^2-M_2^2} \left\{ \left[f_1(x_1) - f_1(x_2)\right] + \frac{1}{6} \left[f_2(x_1)-f_2(x_2)\right] \right\} \, ,
\label{eq:amugeneral}
\end{equation}
where the first term inside the curly brackets corresponds to the chargino contributions, the second term to the neutralino contributions, $x_1 = M_2^2/m_{\widetilde{\nu}}^2$ and $x_2 = \mu^2/m_{\widetilde{\nu}}^2$. In addition,
\begin{equation}
  f_1(x) = \frac{1 - 4 x/3 +x^2/3 + 2 \log(x)/3}{(1-x)^4} \, ,
\end{equation}
and
\begin{equation}
  f_2(x) = \frac{1 - x^2 + 2 x \log(x)}{(1-x)^3} \, .
\end{equation}
In the above we have ignored the small hypercharge induced contributions. It is important to note that for $x \ll 1$, $f_1(x)$ is negative and increases logarithmically in magnitude, $f_1(x) \simeq 1+8 x/3 +2 (1+4x) \log(x)/3$, while$f_2(x)$ tends to one, namely $f_2(x) \to 1 + 2 x (3/2+\log(x))$. On the other hand, in the limit of $x\to 1$, $f_1(x) \to -2/9$ and $f_2(x) \to 1/3$. In general, as stressed above, the lightest chargino contribution is dominant, but the heavier chargino and the neutralino contributions have the opposite sign to the lighter chargino one, providing a significant reduction of the anomalous magnetic moment with respect to the one obtained considering only the lightest chargino contribution. We also note that Eq.~(\ref{eq:amugeneral}) is symmetric under the interchange of $\mu$ and $M_2$, and is indeed valid also in the region in which the second lightest neutralino is Higgsino like, $|M_2| \simgt 2 |\mu| \simgt 4 M_W$, and $m_{\widetilde{\nu}} \simgt |M_2|$.

Let us stress that while the reduction of the SI cross section is obtained for negative value of $\mu \times M_1$, the explanation of the anomalous magnetic moment of the muon demands positive values of $\mu \times M_2$. Hence, a simultaneous explanation of the absence of DM direct detection signals and of the measured value of $a_\mu$ may be naturally obtained for opposite values of the hypercharge and weak gaugino masses, namely $M_2 \times M_1 < 0$.

\section{Flavor Observables}
\label{sec:flav}

As was mentioned in the introduction, $B$-physics observables may be subject to large corrections which should be studied in order to determine the viability of a given low energy supersymmetry scenario~\cite{Gabbiani:1996hi,Altmannshofer:2009ne,Altmannshofer:2012ks}. In the large $\tan\beta$ regime, the most relevant observables are the rate of $(b \to s \gamma)$, $(B^+ \to \tau^+ \nu)$ and $(B_s \to \mu^+\mu^-)$, which have been measured by the BABAR, Belle and LHCb collaborations~\cite{Lees:2012ufa,Belle:2016ufb,Adachi:2012mm,Aaij:2017vad}. The $(b\to s\gamma)$ decay amplitude is affected by a charged Higgs loop contribution~\cite{Hewett:1992is,Misiak:2017bgg}, proportional to $m_t^2/m_{H^\pm}^2$, as well as a stop-chargino loop contribution~\cite{Barbieri:1993av,Ciuchini:1997xe,Degrassi:2000qf,Carena:2000uj}, proportional to 
\begin{equation}
  \mathcal{A}_{\widetilde{t}} \propto \frac{m_t^2 A_t \mu}{m_{\widetilde{t}}^4} \tan\beta \, ,
\end{equation} 
where $A_t$ is the stop mixing parameter and $m_{\widetilde{t}}$ is the characteristic stop mass.

The rate of the decay $(B^+ \to \tau^+ \nu)$ is affected mostly by tree-level charged Higgs contributions~\cite{Hou:1992sy}, which may be large in the region of low charged Higgs masses and large values of $\tan\beta$. The decay amplitude receive corrections that grow like $\tan^2\beta$ (ignoring the loop corrections to the Yukawa couplings), but are suppressed by $1/m_{H^\pm}^2$. As we shall discuss, the LHC is already putting strong constraints on the heavy Higgs masses, which efficiently suppress the corrections to this observable in the region of parameters we shall concentrate on. Finally, the corrections to the amplitude of the decay process $(B_s \to \mu^+\mu^-)$ grow with $\tan^3\beta$, but similar to the corrections to the amplitude for $(b\to s\gamma)$, the loop corrections depend on the precise parameters in the heavy squark sector~\cite{Buras:2002vd}. The stop contributions are proportional to $1/m_{H^\pm}^2$ and are suppressed for small values of $\mu A_t/m_{\tilde{t}}^2$.

In the region of parameters we shall work on, where $\mu \ll m_{\widetilde{t}},m_{\widetilde{b}}$, these flavor observable corrections tend to be small. Moreover, for negative values of $A_t \times \mu$, which we adopt in this work, there are interesting cancellations between the chargino and charged Higgs contributions to $(b \to s \gamma)$. At very large values of $\tan\beta \simeq 60$, however, the corrections proceeding from the stop-chargino contributions may be very large, inducing unacceptable corrections to the $(b\to s \gamma)$ and $(B_s \to \mu^+\mu^-)$ rates for the stop mass parameters we present in our benchmark parameter set in Table~\ref{tab:micromegas_bm}. However, these corrections can be brought under control by pushing the squark masses to larger values while simultaneously reducing the value of $A_t$ to keep consistency with the Higgs mass. Moreover, these loop induced flavor observables may be affected by corrections induced by small, flavor violating gluino couplings~\cite{Gabbiani:1996hi,Carena:2008ue}. Since in this work we are mostly interested in the electroweak sector of the theory, we shall assume a proper value for these flavor observables and we will not expand further on the analysis of the flavor properties of the theory.

\section{Low Energy Supersymmetry Explorations}
\label{sec:Explorations}

As validation for our theoretical expectations outlined above, we explore as an example the MSSM's ability to simultaneously fit the ATLAS three lepton excess, DM relic abundance, and the muon's anomalous magnetic moment while avoiding direct detection constraints. We perform numerical scans using \texttt{micrOMEGAs}~(version 5.0.4)~\cite{Belanger:2001fz,Belanger:2004yn}. To avoid generic bounds from squark and gluino searches, we set their soft masses to 2~TeV. Following the direct detection and $a_\mu$ discussions of Secs.~\ref{sec:DarkMatter} and \ref{sec:g-2}, we require $\mu,\, M_2 < 0$ and $M_1 > 0$, and choose soft slepton masses $M_{\widetilde{L}} \lesssim 500$~GeV. Finally, the SM-like Higgs mass is required to be between 124--126 GeV. Parameters not labeled in the following figures are set to benchmark~(BM) values presented in Table~\ref{tab:micromegas_bm}. 

\begin{figure}[t]
  \centering
  \includegraphics[width=0.75\textwidth]{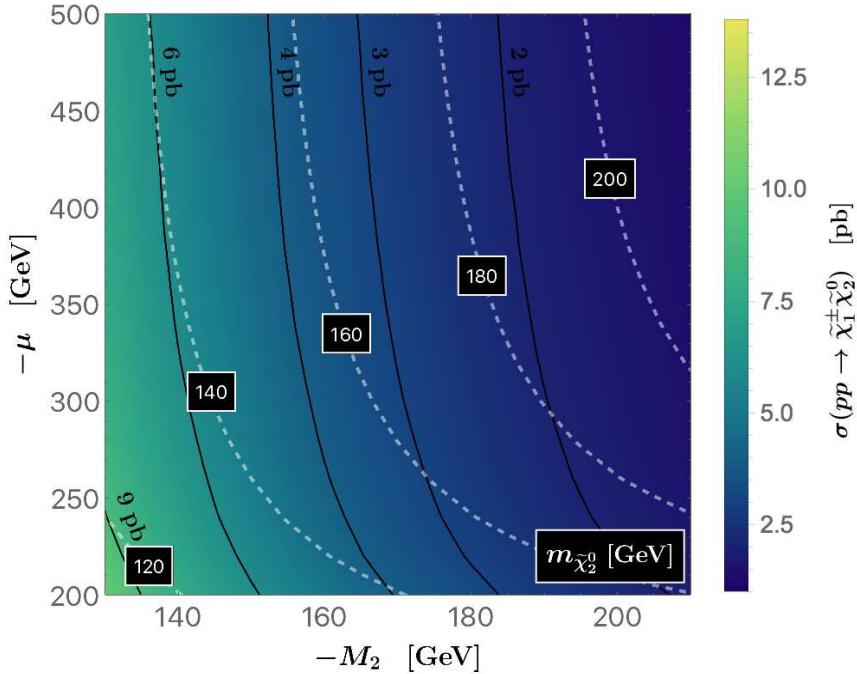}
  \caption{Contours of $\widetilde{\chi}_1^\pm \, \widetilde{\chi}_2^0$ production cross sections (solid black) and $m_{\widetilde{\chi}_2^0}$ (dashed white) in the $\mu$ vs. $M_2$ plane for $\tan\beta$ = 20. All other parameters are fixed to the BM values shown in Table~\ref{tab:micromegas_bm}.}
  \label{fig:atlas_sig}
\end{figure}

We first stress that when considering the LHC production cross section for electroweakinos, unlike the simplified case targeted by the ATLAS and CMS experiments, there can be relevant Higgsino components in $\chi_2^0$ and $\chi_1^\pm$ in the MSSM. This, in general, leads to a reduction of the signal cross section compared to pure wino-like production. To account for this, we calculated the MSSM production cross section to NLO accuracy using \texttt{Prospino2}~\cite{Beenakker:1996ed}. As expected, larger values of $|\mu|$ lead to larger values of the LHC cross section due to the larger wino component of the chargino and second lightest neutralino. This is shown in Fig.~\ref{fig:atlas_sig}, where we present the signal cross sections for the production of the second lightest neutralino in association with the lightest chargino at the LHC in the $M_2$ vs. $\mu$ plane for $\tan \beta = 20$. We note here that since the Higgsino components of a mostly wino-like neutralino are only weakly dependent on $\tan \beta$~\cite{pierce:decomp}, the plot shown will not be modified significantly by varying $\tan \beta$. The mass of the almost degenerate $\widetilde{\chi}_2^0/\widetilde{\chi}_1^\pm$ pair is denoted by the white dashed lines, whereas the color coding shows the values of the LHC production cross section. Black labeled contour lines for the production cross section are also provided to guide the eye. Fig.~\ref{fig:atlas_sig} shows that while the dependence on $\mu$ is mild, there is a strong dependence of the cross section on $M_2$. This is due in part to the fact that in this regime of parameters the chargino mass is predominantly governed by $M_2$. Equally important is the fact that the pure wino cross section is approximately a factor four larger than the pure Higgsino one. This in turn implies that the conventional LHC search bounds on the electroweakino masses become significantly weaker under the inversion of the mass hierarchy between winos and Higgsinos, and in fact a pure Higgsino-like chargino/neutrlino pair is barely constrained by current LHC analyses. This can be seen for example from Fig. 32~(b) in Ref.~\cite{ConvSearch_aux}, by comparing the 95\% C.L. excluded cross sections with the cross sections for the production of pure Higgsino neutralino/chargino pairs~\cite{LHC_XSWG,Fuks:2012qx,Fuks:2013vua}.

As discussed in Sec.~\ref{sec:ATLASexcesses}, in this work we target the signal with the highest significance, which comes from the SR3$\ell_\textrm{ISR}$ search with an estimated signal cross section $\sigma(p p \rightarrow \widetilde{\chi}_1^\pm \widetilde{\chi}_2^0) \sim 3$~pb. Although, as can be seen from Fig.~\ref{fig:sig_xsecs}, this value of the cross section is about $1\sigma$ lower than the central value of the cross section necessary to explain the excess in the signal region SR3$\ell_\textrm{ISR}$, it is in better agreement with the bounds coming from conventional trilepton searches at ATLAS and CMS~\cite{Aaboud:2018jiw, Sirunyan:2018ubx}. From Fig.~\ref{fig:atlas_sig}, it can be seen that a 3~pb signal can be accommodated easily in the MSSM for values of $m_{\widetilde{\chi}_2}^0/m_{\widetilde{\chi}_1}^\pm \sim $ 150-170 GeV. On the other hand, for heavier masses $\sim$ 200~GeV, cross sections of the order of 1.6~pb can be obtained in the MSSM for values of $|\mu|$ of a few hundred GeV. Although such cross sections would only explain two thirds of the tri-lepton event excess found in the RJR analyses, they would be more consistent with the observed excess in the two lepton channel and they would lead to no tension with the results of conventional searches.

Regarding the compatibility of the signal excess in the RJR analysis with existing searches for charginos and neutralinos at the LHC, we reiterate that in Ref.~\cite{Athron:2018vxy} the GAMBIT collaboration analyzed the effects of the production of the relatively light $\widetilde{\chi}_{3,4}^0$, $\widetilde{\chi}_2^\pm$ states generically present in the MSSM in the region of parameters we are investigating. They concluded that the inclusion of these heavier neutralinos and charginos, in combination with the reduced $\widetilde{\chi}_1^\pm \widetilde{\chi}_2^0$ production cross section associated with sizable Higgsino mixing, effectively reduces the tension between the RJR study and previous analyses which set exclusion limits based on a simplified model. It is certainly an intriguing possibility which appears to favor a signal interpretation of the RJR analysis.

\begin{figure}[t]
  \centering
  \begin{tabular}{l l}
    \includegraphics[width=0.38\textwidth]{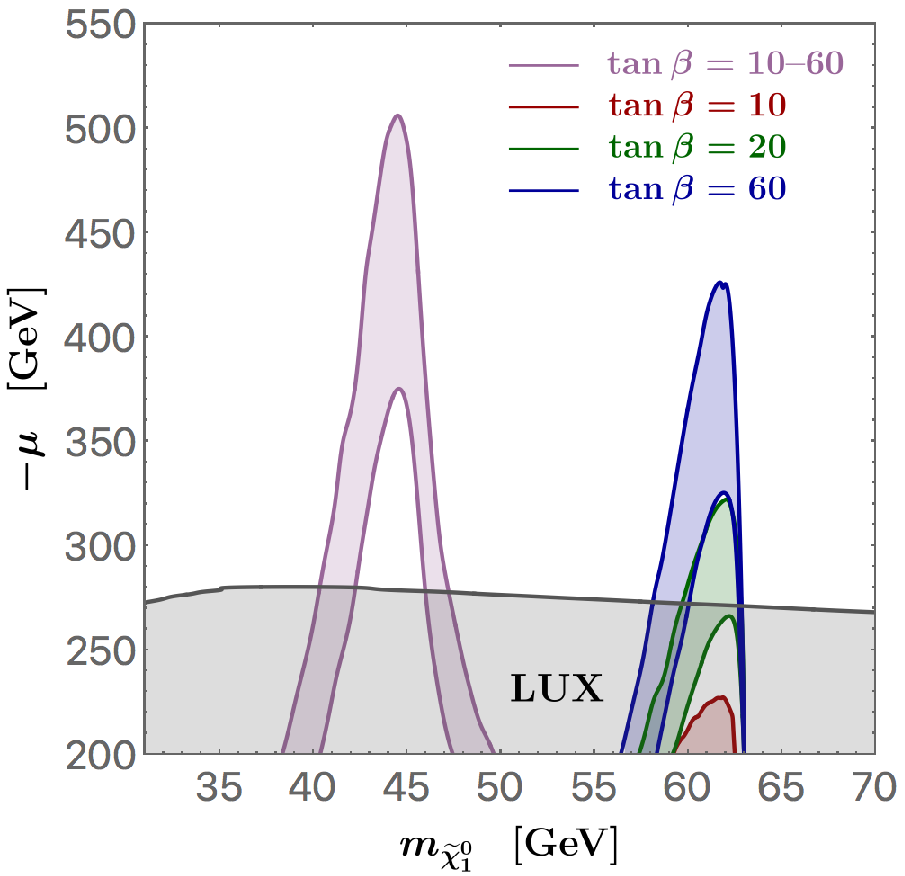}
    &
    \includegraphics[width=0.475\textwidth]{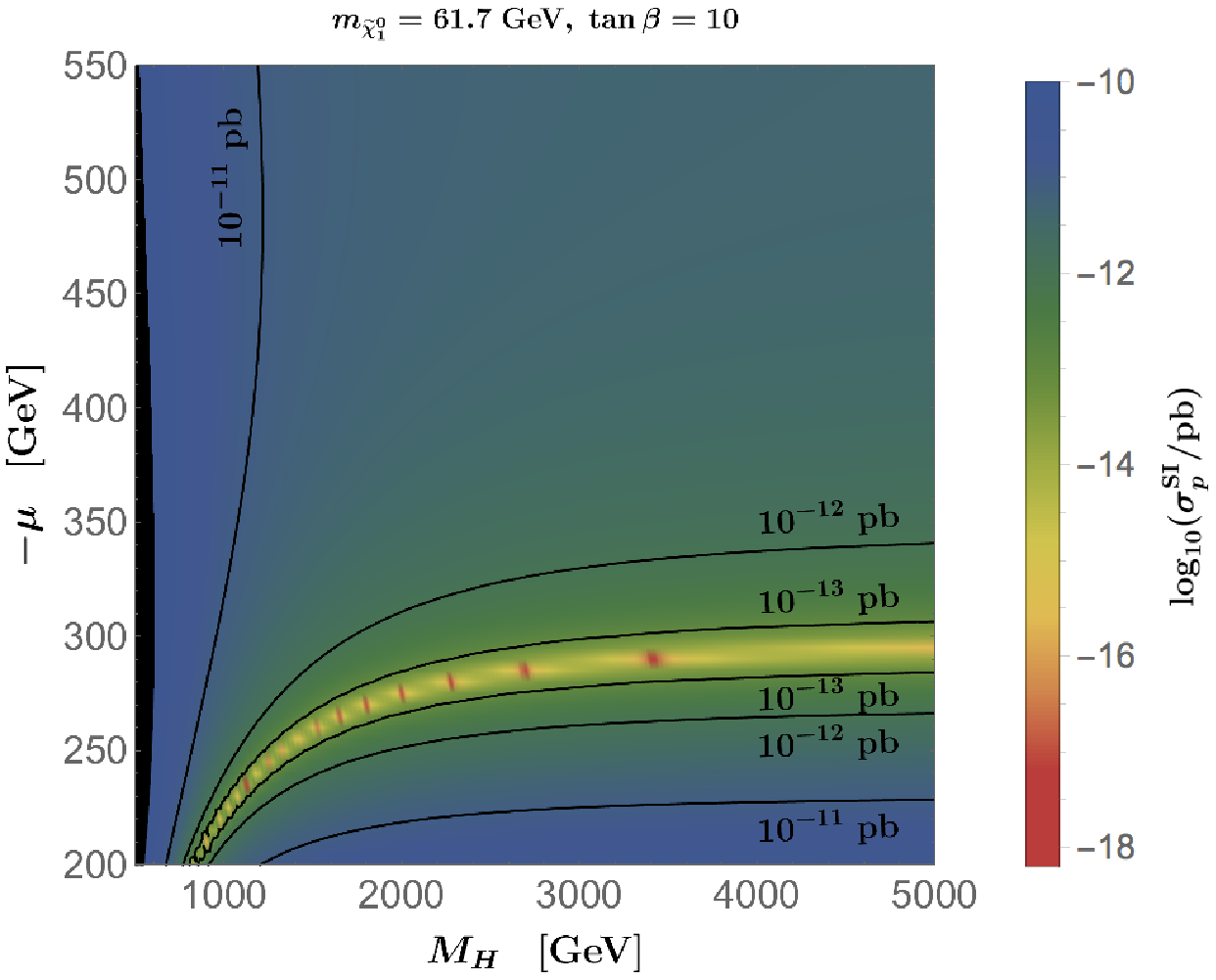}
    \\
    \includegraphics[width=0.475\textwidth]{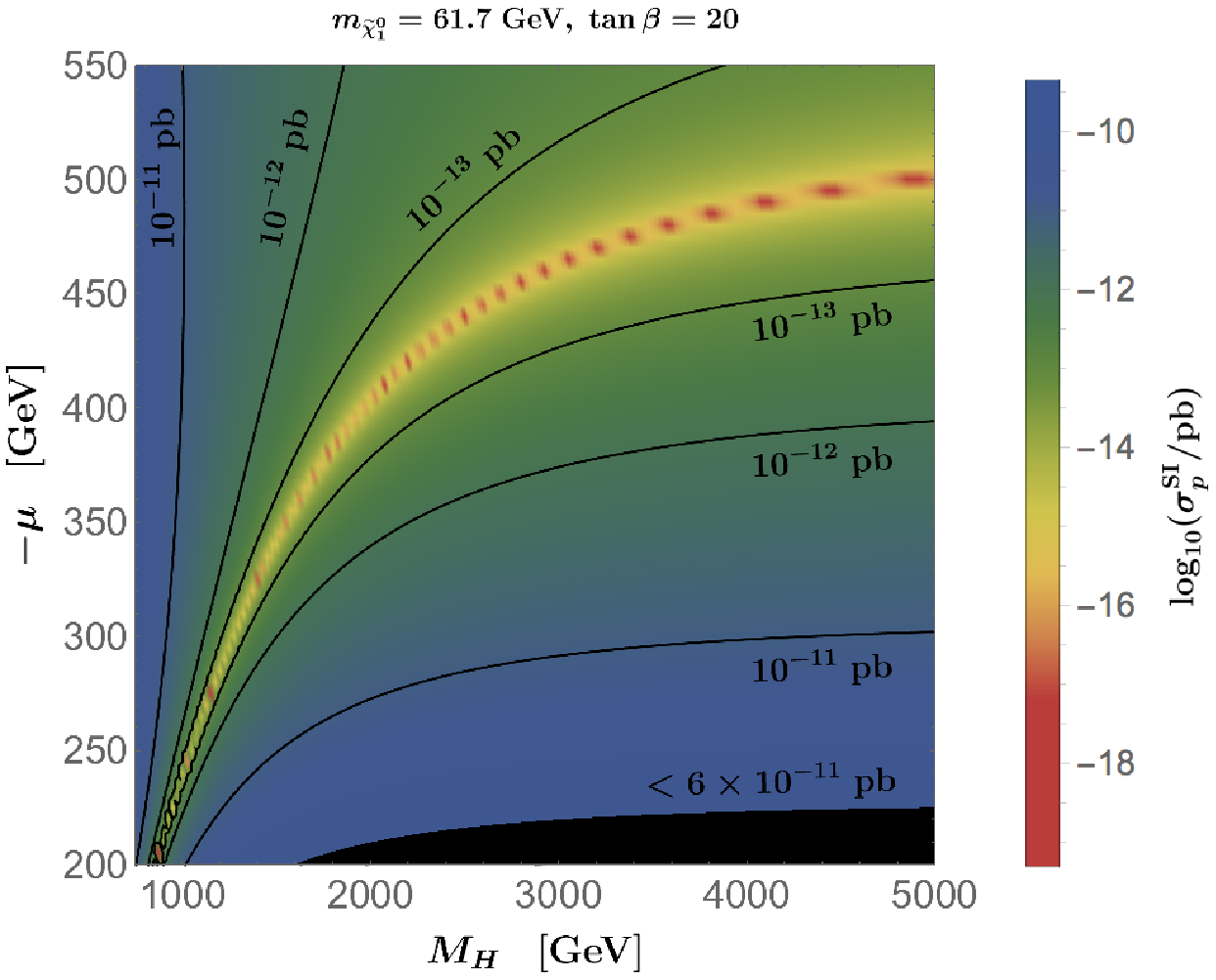}
    &
    \includegraphics[width=0.475\textwidth]{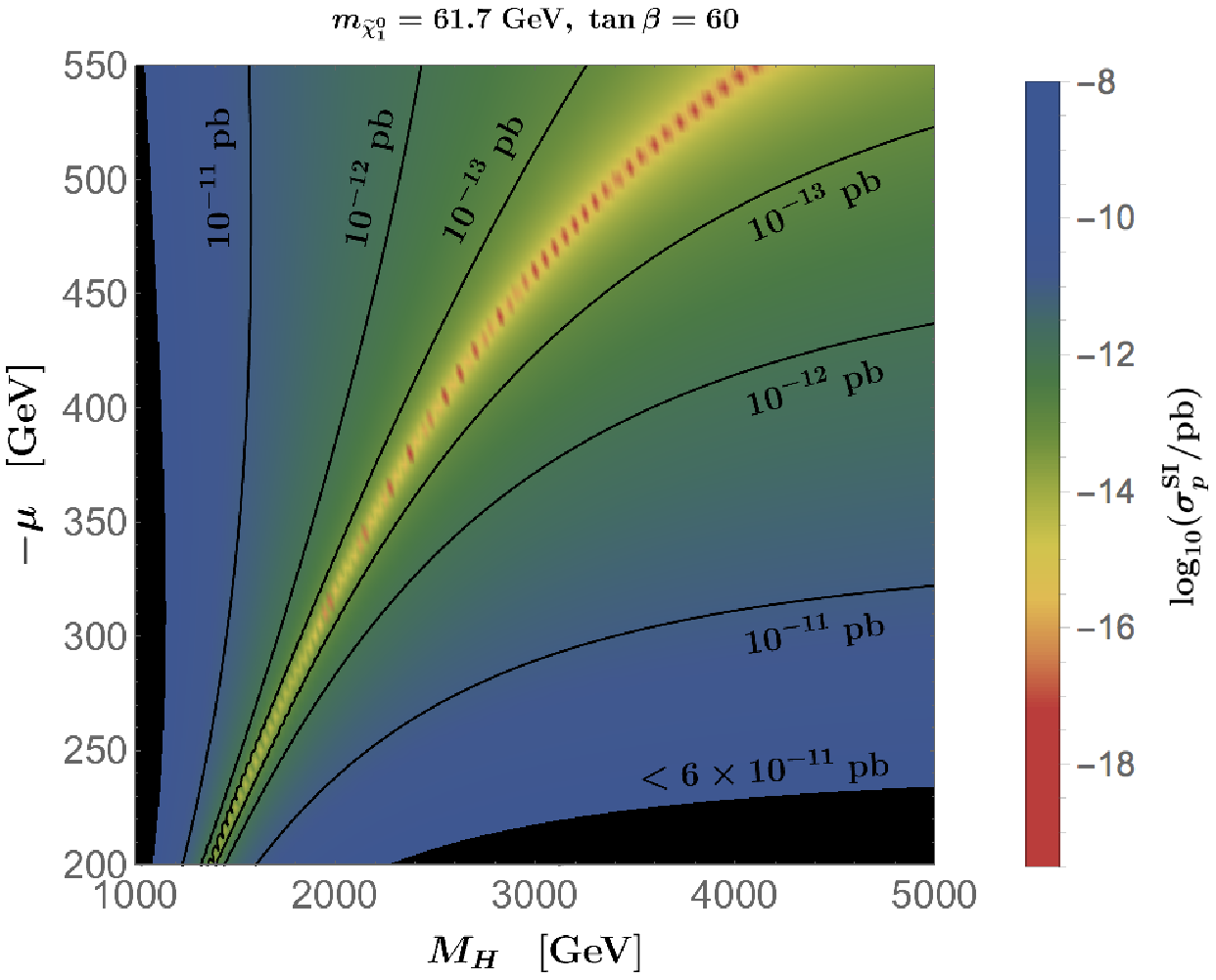}
  \end{tabular}
  \caption{\textit{Top left:} Regions in the $\mu-m_{\widetilde{\chi}_1}$ plane that produce a relic abundance $\Omega_\textrm{CDM} h^2 = 0.12 \pm 50\%$ for different values of $\tan \beta$. The red, green and blue regions correspond to $\tan \beta = 10$, 20, and 60, respectively (corresponding to the Higgs resonance), while the purple region corresponds to the $Z$ resonance which is approximately independent of $\tan \beta$. The lower gray shaded region is excluded by SD constraints set by LUX, which are again approximately independent of the value of $\tan \beta$ for moderate to large values of $\tan \beta$. The three remaining plots show contours of the SI scattering cross section $\sigma_{p}^\textrm{SI}$ in the $M_{H}$--$\mu$ plane for $\tan \beta = 10$ (\textit{top right}), 20 (\textit{bottom left}), and 60 (\textit{bottom right}) with fixed $m_{\widetilde{\chi}_1^0} = 61.7$~GeV. The narrow black regions are excluded by SI constraints set by XENON1T. Other parameters are fixed to the BM values shown in Table~\ref{tab:micromegas_bm}.}
  \label{fig:relic_dd}
\end{figure}

Next we focus on the region of parameters where a phenomenologically viable DM candidate may be obtained consistent with the signal events at the LHC. The upper left panel of Fig.~\ref{fig:relic_dd} shows the region of the $\mu-m_{\widetilde{\chi}_1^0}$ plane that can accommodate the observed relic abundance $\Omega_\textrm{CDM} h^2 = 0.12 \pm 50\%$. As expected, we obtain only the Higgs resonance region~($m_{\widetilde{\chi}_1^0} \sim 60$~GeV) for $\tan \beta = 10$ (red shaded region), $\tan \beta = 20$ (green shaded region), and $\tan \beta = 60$ (blue shaded region), and the $Z$ resonance region~($m_{\widetilde{\chi}_1^0} \sim 45$~GeV) (purple shaded region), which in the large $\tan \beta$ regime is only weakly dependent on $\tan \beta$. We note that the RJR signal is optimized for mass splitting between $m_{\widetilde{\chi}_2^0}/m_{\widetilde{\chi}_1^\pm}$ and $m_{\widetilde{\chi}_1^0} \sim 100$~GeV, such that regardless of the couplings, the decay $(\widetilde{\chi}_2^0 \to h \widetilde{\chi}_1^0)$ is kinematically forbidden, while $(\widetilde{\chi}_2^0 \to Z \widetilde{\chi}_1^0)$ is 100\%. Hence, the Higgs and $Z$ resonance regions would correspond to $m_{\widetilde{\chi}_2^0} / m_{\widetilde{\chi}_1^\pm}\sim$~165 or 145~GeV, respectively. 

As explained in section~\ref{sec:DarkMatter}, for the heavier $m_{\widetilde{\chi}_2^0} / m_{\widetilde{\chi}_1^\pm} \sim$~170--200~GeV, which would prefer $m_{\widetilde{\chi}_1^0} \sim$~70--100~GeV, the only mechanism in the MSSM for obtaining an observationally consistent thermal relic density in the scenario under study would be the $t$-channel interchange of light staus, with masses of the order of the lightest chargino mass. An example of such a scenario would be the addition of $\sim 200$~GeV right-handed staus~\cite{Carena:2012gp,pierce:decomp}. All other sleptons may be kept heavy in order to fulfill the collider and $g-2$ constraints. We have checked that consistency with the relic density and all other phenomenological constraints may be obtained for $\tan\beta \simeq 100$. Such large values of $\tan\beta$ may be acceptable provided there are large corrections to the bottom Yukawa coupling~\cite{Hempfling:1993kv,Hall:1993gn,Carena:1994bv,Guasch:2001wv}, keeping the perturbativity of the bottom sector up to high scales~\cite{Hisano:2010re,Carena:2012mw}. We note that consistent relic density for a heavier slepton spectrum may be also be obtained in the NMSSM, where either co-annihilation with singlinos~\cite{Baum:2017enm} or resonant annihilation through a singlet-like pseudoscalar~\cite{Cheung:2014lqa} may provide the necessary mechanisms without much impact on the collider or direct detection data. 
 
Also shown in the upper left panel of Fig.~\ref{fig:relic_dd} is the region excluded by the SD cross section constraints, which are almost entirely driven by the $Z \widetilde{\chi}_1^0 \widetilde{\chi}_1^0$ coupling and hence depend only weakly on $\tan \beta$~[c.f. Eqs.~(\ref{eq:gBH1}) and~(\ref{eq:SDmu})]. We use the SD limits set by PICO-60 for $\widetilde{\chi}_1^0$--$p$ scattering and LUX for $\widetilde{\chi}_1^0$--$n$ scattering, which are $\sigma_p^\textrm{SD} \approx 4 \times 10^{-5}$~pb~\cite{Amole:2017dex} and $\sigma_n^\textrm{SD} \approx 2 \times 10^{-5}$~pb~\cite{Akerib:2017kat}, respectively, and we verified that the bounds on $\sigma_n^\textrm{SD}$ provide the strongest constraints in our region of parameters. These SD cross section constraints demand the value of $|\mu|$ to be larger than about 270~GeV. Large values of $|\mu|$, on the other hand, are disfavored by the requirement of obtaining the observed relic density. One obtains an upper bound of $|\mu|$ of about 500~GeV, independently of the resonant region, which becomes more stringent for lower values of $\tan \beta$ in the Higgs resonant region. In particular, the combination of the relic density and SD cross section constraints rules out values of $\tan\beta < 10$ in the Higgs resonant annihilation region.

The remaining panels show the SI direct detection cross section for $m_{\widetilde{\chi}_1^0}=61.7$~GeV in the $\mu-m_H$ plane for $\tan \beta = 10$ (upper right), 20 (lower left), and 60 (lower right), where the regions excluded by current XENON1T results are denoted in black. We use the current SI limit set by XENON1T, $\sigma^\textrm{SI} \simlt 6 \times 10^{-11}$~pb~\cite{Aprile:2018dbl}. We see that in this region of parameters the SI cross section tends to be naturally smaller than the current experimental limit. The behavior of the SI cross section may be easily understood by the approximate formulae, Eqs.~(\ref{eq:siglargetb}) and (\ref{eq:blindspot}). The yellow and red bands show the presence of blind spots, where the SI cross section is significantly reduced and may be below the neutrino floor and beyond the reach of near future experiments. Beyond these regions, the SI cross sections may be probed by the current and near future experiments, but they rarely exceed $10^{-11}$~pb. Although one would naturally expect larger cross sections for larger values of $\tan \beta$, the presence of blind spots tends to suppress the SI cross section to values below $10^{-11}$~pb even for $\tan \beta = 60$, unless the heavy Higgs bosons take values significantly larger than or below the TeV~scale. We note that the results shown in these plots are only weakly dependent on the value of $m_{\widetilde{\chi}_1}$, as long as $m_{\widetilde{\chi}_1}^2 \ll \mu^2$. Whereas the precise location of the blind-spot in the $\mu-m_H$ plane does depend on $m_{\widetilde{\chi}_1}$, the qualitative behavior of the SI cross section discussed remains the same. 

Let us stress that heavy Higgs masses below a~TeV are currently disfavored by searches for heavy resonances decaying to tau lepton pairs~\cite{Aaboud:2017sjh,Sirunyan:2018zut}. These constraints, however, may be avoided for $\tan\beta\simlt 20$. Indeed, for scenarios with light electroweakinos like the one we are analyzing, the limit on the heavy Higgs mass is lower. For $\tan\beta = 10$, due to the relatively small coupling of the $\tau$ lepton to the heavy Higgs bosons, it becomes of the order of $|M_2| + |\mu|$, but becomes stronger, increasing to about 900~GeV, for values of $\tan\beta = 20$~\cite{Bahl:2018zmf}. At $\tan\beta = 60$, the bound is about 1.5~TeV, excluding the region with large SI cross sections at the left of the blind-spot band in Fig.~\ref{fig:relic_dd} at this values of $\tan\beta$. Moreover, precision measurements of the properties of the SM-like Higgs demand the heavy Higgs bosons to be heavier than about 600~GeV in this regime. Hence, if the small excess observed by XENON1T~\cite{Aprile:2018dbl} were a signal of the presence of DM, within this scenario it would lead to the preference for Higgs masses of the order of 600~GeV--1~TeV and $10 \simlt \tan\beta \simlt 20$ (with larger values of the heavy Higgs mass for larger values of $\tan\beta$), or for Higgs masses $\gtrsim$ 2 TeV for $\tan\beta = 60$. As shown in the left-hand upper panel of Fig.~\ref{fig:relic_dd} this would lead to a preference for the $Z$-resonance annihilation region for the smaller $\tan\beta$ values. 

\begin{figure}[t]
  \centering
  \includegraphics[width=0.47\textwidth]{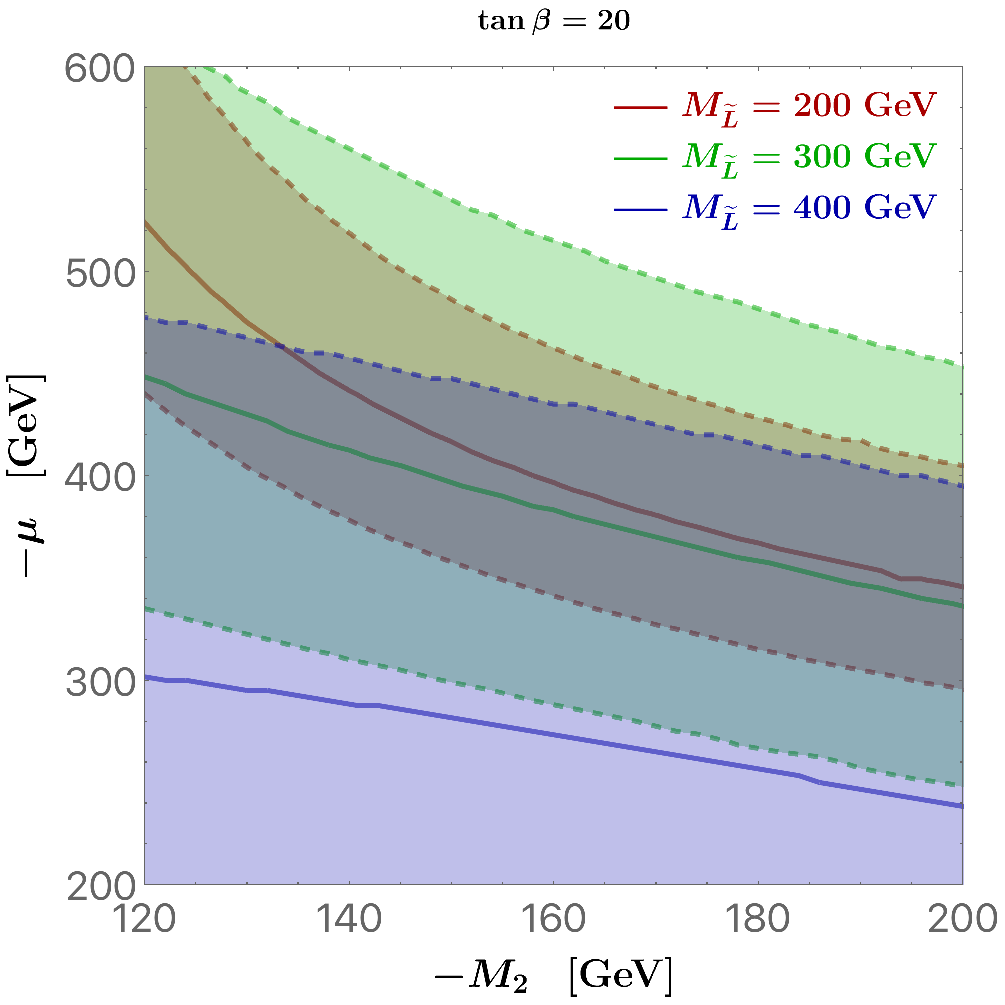}
  \hspace*{0.04\textwidth}
  \includegraphics[width=0.45\textwidth]{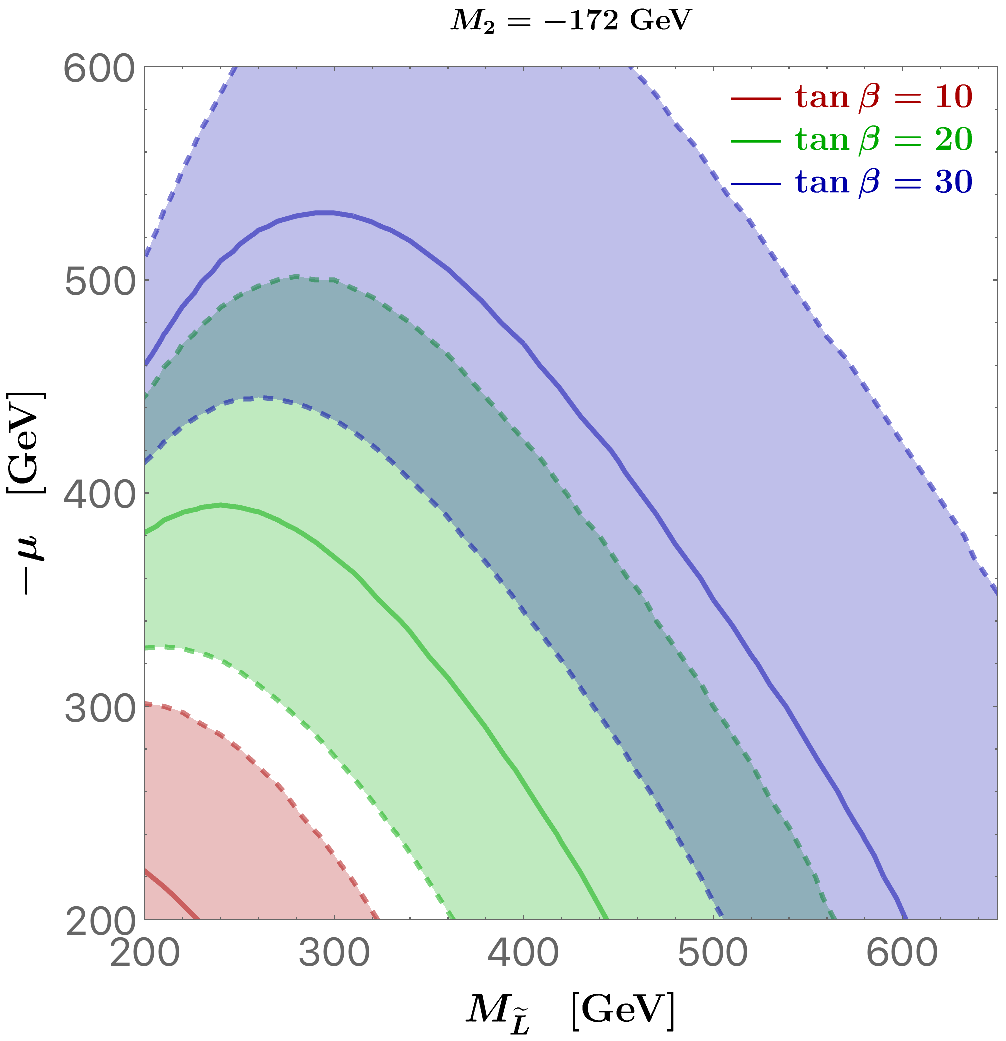}
  \caption{Regions of parameter space that produce the observed excess in the anomalous magnetic moment of the muon. Solid lines denote consistency with the current experimental values, while shaded regions show $1\sigma$ variations. \textit{Left:} The $M_2$ and $\mu$ dependence for several choices of the slepton soft mass parameter $M_{\widetilde{L}}$ and $\tan \beta = 20$. \textit{Right:} The $M_{\widetilde{L}}$ and $\mu$ dependence for several values of $\tan \beta$ and $M_2 = -172$~GeV. Other parameters not shown are fixed to the BM values shown in Table~\ref{tab:micromegas_bm}.}
  \label{fig:gminus2}
\end{figure}

Finally, Fig.~\ref{fig:gminus2} shows the region of parameter space that accommodates the observed deviation of the anomalous magnetic moment of the muon with respect to the SM prediction. The left panel of Fig.~\ref{fig:gminus2} shows the preferred values of $M_2$ and $\mu$ for different values of the slepton masses, and $\tan\beta = 20$. For simplicity, we have assumed equal soft supersymmetry breaking parameters for left- and right-handed sleptons, characterized by $M_{\widetilde{L}} \simeq M_{\widetilde \nu} $. The solid lines denote the values of $\mu$ leading to agreement with the observed value of $a_\mu$, while the shaded bands show the range of $\mu$ consistent with the current $1\sigma$ experimental uncertainty on this quantity. Overall, the dependence of $a_\mu$ on the supersymmetry breaking mass parameters is in agreement with our general expectations based on Eqs.~(\ref{eq:amu}) and (\ref{eq:amugeneral}). Lighter (heavier) sleptons imply larger (smaller) preferred values of $|\mu|$, with values of $|\mu|$ in the range 200--500~GeV for this value of $\tan\beta$ and slepton masses at the weak scale. 
 
The right panel of Fig.~\ref{fig:gminus2} shows the dependence of the preferred values of the slepton masses and the Higgsino mass parameter for different values of $\tan\beta$ and $M_2=-172$ GeV. While values of $\tan\beta = 10$ demand values of these parameters of the order of 200-300 GeV, the slepton masses can be significantly larger for values of $\tan\beta = 60$. In particular, for $\tan\beta = 60$ and $|\mu| = 300$~GeV, slepton masses of the order of 500~GeV (700~GeV) are consistent with the central experimental value (a deviation of one standard deviation with respect to the central value).

Let us comment that as can be seen from the right panel of Fig.~\ref{fig:gminus2}, for small values of the slepton masses, $M_{\widetilde{L}} < |\mu|$, which are not described by Eq. (\ref{eq:amugeneral}), there is a turning point in the contours of constant $a_\mu$, which tends to lower values of $|\mu|$. This is induced by an increase of the contribution of neutralinos compared to the one of charginos. Also, the right-handed slepton contribution become relevant in this regime. Such light right-handed sleptons, however, are being constrained by the LHC, which is putting relevant bounds on slepton masses~\cite{Aaboud:2017leg,Aaboud:2018jiw}. For instance, the bound on degenerate first and second generation left and right handed sleptons decaying into leptons and missing energy is about 520~GeV. In our setup, however, only the right-handed sleptons decay directly into leptons and missing energy. The left-handed sleptons instead decay first into chargino and second-lightest neutralino states, which as we discussed before, decay into weak gauge bosons and missing energy. Hence, the bounds on these sleptons are expected to be significantly weaker than the ones associated with the decay into just leptons and missing energy. Regarding the limit on the right-handed sleptons, the collective cross section of first and second generation sleptons with mass of about 520~GeV is about 1~fb, while the one of 400~GeV right-handed sleptons is also about 1~fb and hence at the edge of the LHC limit. However, since the right-handed sleptons do not play an important role in determining $a_\mu$, it is enough to make them a few tens of the GeV heavier to easily avoid the current LHC limits, without affecting any of the essential features of this scenario.

As a concrete example, we present a BM parameter set satisfying all of the constraints discussed above. The MSSM parameters are shown in Table~\ref{tab:micromegas_bm} for $\tan \beta = 20$, and the associated mass spectrum (generated with \texttt{SuSpect2}~\cite{Djouadi:2002ze}, including radiative corrections) is shown in Table~\ref{tab:suspect_bm}. The NLO production cross section in the MSSM corresponding to our BM masses is
\begin{equation}
  \sigma(p p \rightarrow \chi_1^\pm \chi_2^0) = 2.92~\textrm{pb} \, ,
\end{equation} 
for the sum of $\widetilde{\chi}_1^+ \widetilde{\chi}_2^0$ and $\widetilde{\chi}_1^- \widetilde{\chi}_2^0$ production. The lightest neutralino annihilates via the Higgs resonance, giving a relic abundance of
\begin{equation}
  \Omega_\textrm{CDM} h^2 = 0.121 \, ,
\end{equation}
while the cross sections for SI and SD direct detection are
\begin{align}
  \begin{array}{l c l}
    \sigma_p^\textrm{SI} = 6.82 \times 10^{-13}~\textrm{pb} \, , & \phantom{abcde} & \sigma_p^\textrm{SD} = 1.70 \times 10^{-5}~\textrm{pb} \, , \\
    \sigma_n^\textrm{SI} = 4.70 \times 10^{-13}~\textrm{pb} \, , & & \sigma_n^\textrm{SD} = 1.33 \times 10^{-5}~\textrm{pb} \, .
  \end{array}
\end{align}
Finally, the MSSM contribution to the muon's anomalous magnetic moment is estimated to be
\begin{equation}
  a_\mu^\textrm{MSSM} = 248 \times 10^{-11} \, .
\end{equation}

\begin{table}[t]
  \centering
  \renewcommand{\arraystretch}{1.2}
  \begin{tabular}{l c | l c | l c | l c} 
    \hline
    Param. & [GeV] & Param. & [GeV] & Param. & [GeV] & Param. & [GeV] \\
    \hline \hline
    $\mu$ & -300 & $M_2$ & -172 & $M_{\widetilde{L}}$ & 400 & $M_{H}$ & 1500 \\
    $M_1$ & 63.5 & $M_3$ & 2000 & $M_{\widetilde{Q}}$ & 2000 & $A_t$ & 3000 \\
    \hline
  \end{tabular}
  \caption{Benchmark values of MSSM input parameters for \texttt{micrOMEGAs} with $\tan \beta = 20$. The squark and slepton soft masses are degenerate between generations and chiralities.}
  \label{tab:micromegas_bm}
\end{table}

\begin{table}[t]
  \centering
  \renewcommand{\arraystretch}{1.2}
  \begin{tabular}{l c | l c | l c | l c}
    \hline
    Part. & $m$ [GeV] & Part. & $m$ [GeV] & Part. & $m$ [GeV] & Part. & $m$ [GeV] \\
    \hline \hline
    $h$ & 125.84 & $\widetilde{\chi}_1^\pm$ & 165.0 & $\widetilde{\nu}_e$ & 395.0 & $\widetilde{u}_R$ & 2069.8 \\
    $H$ & 1500.03 & $\widetilde{\chi}_2^\pm$ & 333.6 & $\widetilde{\nu}_\mu$ & 395.0 & $\widetilde{u}_L$ & 2069.5 \\
    $H_3$ & 1500.00 & $\widetilde{\tau}_1$ & 389.5 & $\widetilde{\nu}_\tau$ & 395.0 & $\widetilde{d}_R$ & 2070.3 \\
    $H^\pm$ & 1502.38 & $\widetilde{\tau}_2$ & 415.0 & $\widetilde{g}$ & 2129.2 & $\widetilde{d}_L$ & 2071.0 \\
    $\widetilde{\chi}_1^0$ & 61.7 & $\widetilde{e}_R$ & 402.4 & $\widetilde{t}_1$ & 1927.7 & $\widetilde{s}_R$ & 2070.3 \\
    $\widetilde{\chi}_2^0$ & 164.8 & $\widetilde{e}_L$ & 402.6 & $\widetilde{t}_2$ & 2131.6 & $\widetilde{s}_L$ & 2071.0 \\
    $\widetilde{\chi}_3^0$ & 314.2 & $\widetilde{\mu}_R$ & 402.4 & $\widetilde{b}_1$ & 2067.1 & $\widetilde{c}_R$ & 2069.8 \\
    $\widetilde{\chi}_4^0$ & 331.2 & $\widetilde{\mu}_L$ & 402.6 & $\widetilde{b}_2$ & 2074.1 & $\widetilde{c}_L$ & 2069.5 \\
    \hline
  \end{tabular}
  \caption{Benchmark mass spectrum generated from the input parameters of Table~\ref{tab:micromegas_bm}.}
  \label{tab:suspect_bm}
\end{table}

The production cross section required to accommodate the central value excesses in the three lepton searches at ATLAS for $(m_{\widetilde{\chi}_1^\pm/\widetilde{\chi}_2^0},\, m_{\widetilde{\chi}_1^0}) = (165,\, 61.7)$~GeV is approximately 4~pb~(c.f. Fig.~\ref{fig:sig_xsecs}). While our BM cross section remains $\sim 1\sigma$ below this central value, we again stress that this may alleviate some tension with previous analyses. We note that lower values of $m_{\widetilde{\chi}_1^\pm/\widetilde{\chi}_2^0}$, as preferred for $\widetilde{\chi}_1^0$ resonant annihilation to the $Z$ boson, generally improve the consistency with the trilepton RJR searches at the expense of increasing the tension with previous analyses. Regarding the direct detection cross sections for our BM point, while they are sufficiently suppressed to evade current limits, they may be probable in the near future through SD interactions. Lastly, we see that the resulting value of $a_\mu$ is well within $1\sigma$ of the currently observed experimental value. 

Finally, we would like to reiterate that the excess of events observed in the ATLAS RJR analysis is interesting but cannot be yet taken as a significant signal of new physics. We present this BM point only as an example of the possible parameters in the electroweak sector consistent with current data. Quite generally, we show that if future LHC data provides a confirmation of electroweakinos at the weak scale, it is not difficult to fulfill other observational and experimental constraints as well. Accommodating the observed relic density is generically the most stringent requirement.

\section{Conclusions}
\label{sec:Conclusions}

Despite a lack of any conclusive evidence for its presence at the weak scale, supersymmetry remains a well motivated extension of the SM, and may answer many open questions in particle physics. In this article we have presented a study of the current constraints on the electroweak sector in low energy supersymmetry models. As an example, we have taken gaugino and Higgsino mass parameters that can be consistent with a new physics interpretation of recent event excesses in the ATLAS search for electroweakinos using the RJR method. The large cross sections associated with these excesses imply that the second lightest supersymmetric particle must be light, with a mass below about 200~GeV and with a large wino component. 

Overall, collider and DM relic density constraints lead to masses for the lightest chargino and neutralinos of 150~GeV~$ \simlt m_{\widetilde{\chi}_1^\pm,\widetilde{\chi}_2^0} \simlt$~200~GeV and 45~GeV~$\simlt m_{\widetilde{\chi}_1^0} \simlt$~100~GeV, respectively. The lower range of masses for the chargino and second lightest neutralino, of about 150~GeV, and of the lightest neutralino, of order 50~GeV, leads to consistency with the most significant events excesses, associated with trileptons plus missing energy. Moreover, for this range of masses the observed relic density may be obtained through the resonant annihilation of the lightest neutralino via either the SM-like Higgs or the $Z$-gauge boson. On the other hand, the higher chargino mass range, of about 200~GeV, and lightest neutralino masses, of about 100~GeV, allows for a better description of the dilepton plus missing energy events, but at the price of worsening the description of the trilepton plus missing energy ones. Barring non-thermal mechanisms, the only way of obtaining the observed relic density for heavy squarks and the lightest neutralino in this range of masses in the MSSM is through the interchange of light staus, with mass close to~200~GeV and large values of $\tan\beta\simeq 100$. This may be also obtained in simple extensions of the MSSM, like the NMSSM, without affecting the phenomenological signatures we analyze in this work.

Consistency with the observed relic density and the current bound on the SD cross section requires values of 500~GeV~$\simgt |\mu| \simgt 270$~GeV. The smaller values of $|\mu| \simeq$~300~GeV also lead to lower LHC production cross sections, reducing the tension with the bounds coming from conventional searches, while allowing $1\sigma$ consistency with the RJR analysis. The heavy Higgs sector remains at energy scales of the order of 1~TeV and provides a reduction of the SI cross section rate, which is naturally below the current bounds on this quantity. Values of the SI cross section close to the current experimental bound may be obtained for values of $10 \simlt \tan\beta \simlt 20$ and 500~GeV~$\simlt M_H \simlt$~1TeV, which may also be probed at the LHC through heavy Higgs decays to di-tau searches in the near future. Moreover, values of $\tan\beta \simgt 10$ and left-handed sleptons with masses of the order of 200--500~GeV are required to explain the observed anomalous magnetic moment of the muon. Negative values of $\mu \times M_1$ and positive values of $\mu \times M_2$ are able to accommodate the current constraints on the SI cross section and the observed anomalous magnetic moment of the muon, respectively. Squarks and, in particular, gluinos, remain significantly heavier than the weakly interacting particles. This, together with the information on the weak gaugino masses discussed above, favors schemes with highly non-universal gaugino mass parameters. 

We stress that most of the conclusions of this work remain valid even if the LHC production cross sections of $\widetilde{\chi}_1^\pm \widetilde{\chi}_2^0$ are lower than the one suggested by the ATLAS RJR analysis. For a given chargino mass and value of $|\mu|$, such lower cross sections would be associated with larger values of $|M_2|$. The behavior of the relevant observables may be understood by noticing that in the relevant region of parameters the $g-2$ results, Eq.~(\ref{eq:amugeneral}), are invariant under the interchange of $M_2$ and $\mu$. Moreover, for a given value of $M_1$, the DM relic density and its SI and SD interaction cross sections depend only on the value of $\mu$. Hence, all the results presented here are easily extrapolated to the case of larger $|M_2|$ and can be understood from the analytical expressions and numerical results shown in this article.

In summary, this work has shown that the electroweak sector of low energy supersymmetry models can be naturally compatible with the observed DM relic density, the absence of direct DM detection signals, and the observed value of $a_\mu$. Combined with a new physics interpretation of the recent event excesses reported by the ATLAS collaboration in the RJR analysis, this consequently leads to a consistent picture in which the characteristic mass parameters of the gaugino, Higgsino, heavy Higgs, and slepton sectors may be determined. The study we have performed is useful to understand the behavior of DM and ($g_\mu-2$) observables for electroweakino masses of a few hundred GeV, even if these ATLAS event excesses were not confirmed. The analysis of the complete run II LHC data set, together with improvements in the determination of the range of SI and SD cross sections and the anomalous magnetic moment of the muon will further probe this attractive, beyond the standard model physics scenario in the near future.

\section*{\sc Acknowledgments}

This manuscript has been authored by Fermi Research Alliance, LLC under Contract No. DE-AC02-07CH11359 with the U.S. Department of Energy, Office of Science, Office of High Energy Physics. The United States Government retains and the publisher, by accepting the article for publication, acknowledges that the United States Government retains a non-exclusive, paid-up, irrevocable, world-wide license to publish or reproduce the published form of this manuscript, or allow others to do so, for United States Government purposes. Work at University of Chicago is supported in part by U.S. Department of Energy grant number DE-FG02-13ER41958. Work at ANL is supported in part by the U.S. Department of Energy under Contract No. DE-AC02-06CH11357. NRS and JO are supported by Wayne State University and by the U.S. Department of Energy under Contract No. DESC0007983. MC, NRS and CW would like to thank the Aspen Center for Physics, which is supported by National Science Foundation grant PHY-1607611, for the kind hospitality during the completion of this work. We would like to thank Y. Bai, D. Cinabro, T. Effert, H. Gray, R. Harr, P. Huang, P. Jackson, D. Liu, J. Liu, D. Miller, A. Petridis and X. Wang for useful discussions and comments.

\bibliography{mybibfile}
\bibliographystyle{JHEP}
 
\end{document}